\documentclass[final,1p,times]{elsarticle} 
\usepackage{graphicx} 
\usepackage{amssymb} 
\usepackage{amsthm} 
\usepackage{lineno} 

\journal{Nuclear Physics A} 
\begin{document} 

\begin{frontmatter} 


\title{A transport calculation with an embedded (3+1)d hydrodynamic evolution: Elliptic flow as a function of transverse momentum at SPS energies}

\author{H.~Petersen$^a$, J.~Steinheimer$^a$, G.~Burau$^{a,b}$ and M.~Bleicher$^a$}

\address[a]{Institut f\"ur Theoretische Physik, Johann Wolfgang Goethe-Universit\"at, Max-von-Laue-Str.~1, D-60438 Frankfurt am Main, Germany}
\address[b]{Frankfurt Institute for Advanced Studies (FIAS), Ruth-Moufang-Str.~1, D-60438 Frankfurt am Main, Germany}

\begin{abstract} 
The transverse momentum dependence of elliptic flow of pions calculated in a full (3+1)d Boltzmann approach with an intermediate hydrodynamic stage for heavy ion reactions for CERN-SPS energies is discussed in the context of the experimental data. At higher SPS energies, where the pure transport calculation cannot account for the high elliptic flow values, the smaller mean free path in the hydrodynamic evolution leads to larger elliptic flow. Due to the more realistic initial conditions and the incorporated hadronic rescattering the results are in line with the experimental data. Within this integrated dynamical approach different equations of state are applied without adjusting the initial state and the freeze-out conditions. We employ a hadron gas equation of state to investigate the differences in the dynamics and viscosity effects, a chiral equation of state with a moderate first order phase transition and a critical endpoint and a bag model equation of state with a large latent heat. The elliptic flow results are shown to be rather insensitive to changes in the equation of state.  
\end{abstract} 

\end{frontmatter} 


\section{Introduction}
\label{intro}
Transverse collective flow has early been predicted as an observable to probe heated and compressed nuclear matter \cite{Stoecker:1986ci,Voloshin:2008dg}. Elliptic flow, the anisotropy parameter that quantifies the momentum space anisotropy in the transverse plane of the outgoing particles of a heavy ion reaction, is a result of the asymmetric pressure gradients that are present in the course of the evolution. Therefore, it is expected to be sensitive to the equation of state (EoS) and has been proposed as a signal for the QGP phase transition \cite{Stoecker:2004qu,Stoecker:2004xc}. The initial conditions and the mean free path during the high energy density stage of the evolution are crucial for the development of the momentum anisotropy that is measured as $v_2$. 

The consistent dynamical modeling of heavy ion reactions is essential to extract information about the created matter from the final state particle distributions. During the last years, various so called microscopic plus macroscopic hybrid approaches have been developed and shown to be very successful in describing many observables at the same time \cite{Bass:2000ib,Teaney:2000cw,Grassi:2005pm,Hirano:2005xf,Nonaka:2006yn}. These hybrid approaches combine the advantages of hydrodynamics, and non-equilibrium hadronic transport approaches. Here we use the same technique and employ a transport approach with an embedded three-dimensional ideal relativistic one fluid evolution for the hot and dense stage of the reaction based on the Ultra-relativistic Quantum Molecular Dynamics (UrQMD) model \cite{Steinheimer:2007iy,Petersen:2008dd,Petersen:1900zz,Petersen:2009vx}. 

\section{The Hybrid Approach}
\label{model}

UrQMD \cite{Bass:1998ca,Bleicher:1999xi,Petersen:2008kb} is a hadronic transport approach which simulates multiple interactions of ingoing and newly produced particles, the excitation and fragmentation of color strings and the formation and decay of hadronic resonances. The UrQMD model is used to account for the non-equilibrium dynamics in the initial stage and the final state rescatterings during the heavy ion reaction. 

\begin{figure}[ht]
\vspace{-0.5cm}
\centering  
\includegraphics[width=7cm,height=5cm]{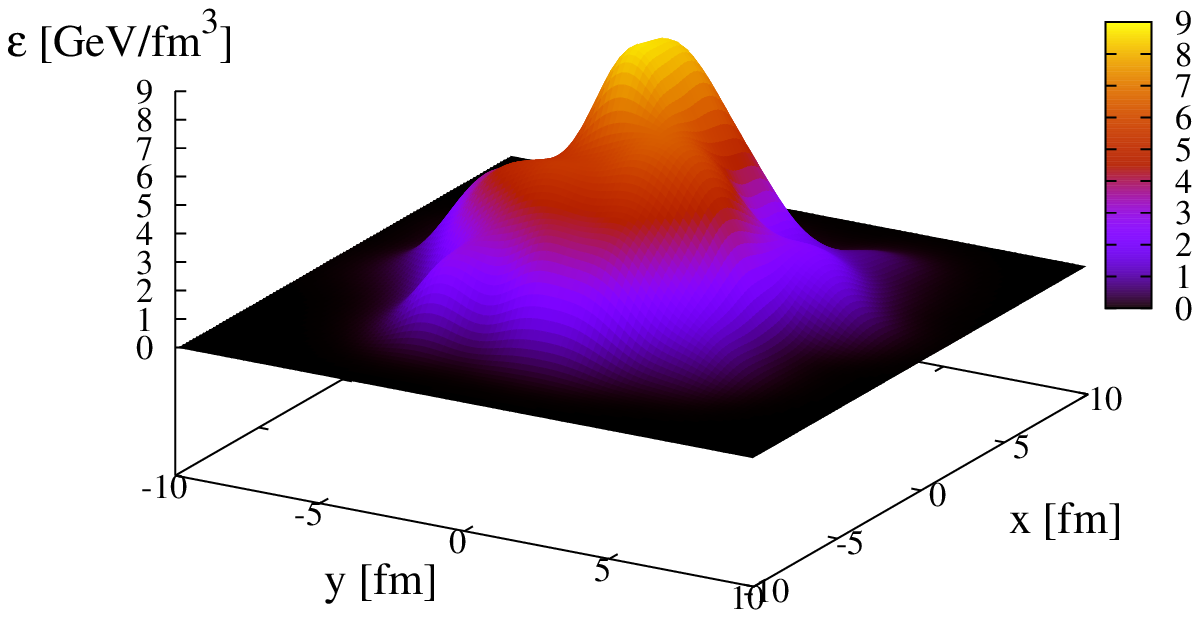}
\includegraphics[width=0.45\textwidth]{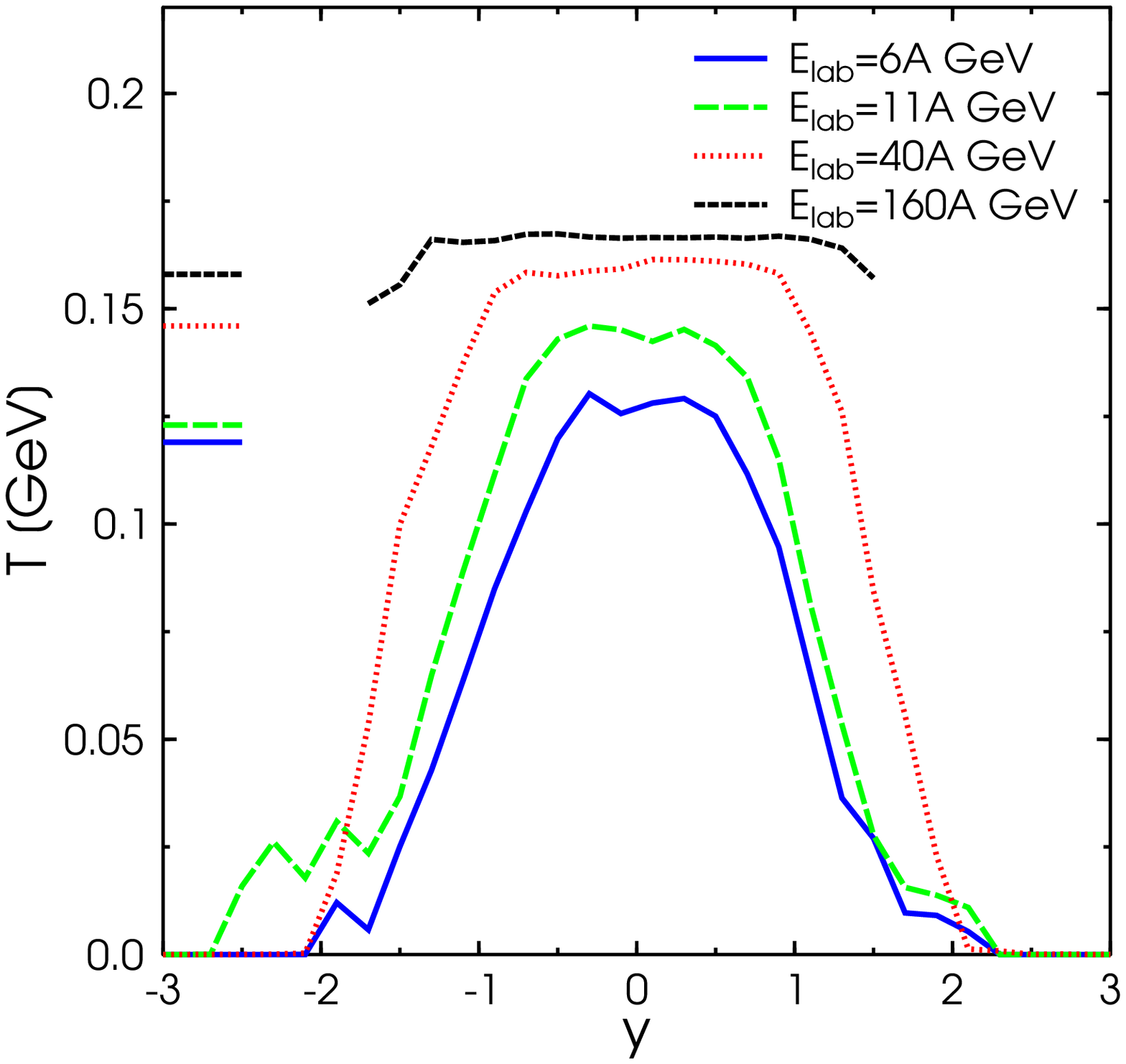}
\vspace{-0.5cm}
\caption{Left: Energy density distribution in the $x-y$-plane for one central ($b=0$ fm) Pb+Pb collision at $E_{\rm lab}=40A~$GeV. Right: Rapidity profile of the transition temperatures in the spatial plane with $x=y=0$ fm for central Au+Au/Pb+Pb collisions at four different beam energies ($E_{\rm lab}=6,~11,~40$ and $160A~$GeV). The small lines on the temperature axis indicate the corresponding chemical freeze-out temperature as extracted from \cite{Cleymans:2006qe}.}
\label{edens}      
\end{figure}

The hydrodynamic fields are initialized after the two nuclei have passed through each other \cite{Steinheimer:2007iy}. The spectators continue to propagate in the cascade during the whole hydrodynamic evolution. As an example, Fig. \ref{edens} (left) shows the energy density distribution in the transverse plane for one single central Pb+Pb collision at $E_{\rm lab}=40A$ GeV. The distribution is not symmetric in any direction and fluctuates event-by-event. 

Starting from these initial conditions a full (3+1) dimensional ideal hydrodynamic evolution is performed using the SHASTA algorithm \cite{Rischke:1995ir,Rischke:1995mt}. If the energy density $\varepsilon$ drops below five times the ground state energy density (i.e. $\sim 730 {\rm MeV/fm}^3$) in all cells of a transverse slice of thickness  $\Delta z = 0.2 $fm, the hydrodynamic fields in this given slice are transformed to particle degrees of freedom via the Cooper-Frye equation. This procedure mimics an approximate iso-eigentime transition (see \cite{Li:2008qm} for more details). The created particles proceed in their evolution in the hadronic cascade (UrQMD) where rescatterings and final decays are calculated until all interactions cease and the system decouples. 

By doing this we obtain a rapidity independent transition temperature even for the highest beam energies (see Fig. \ref{edens} (right)). The midrapidity transition temperatures are slightly larger than the chemical freeze-out temperatures that have been parametrized by Cleymans et al. \cite{Cleymans:2006qe} because in our case the chemical freeze-out happens during the following cascade evolution.

Serving as an input for the hydrodynamical calculation the EoS should strongly influence the dynamics of an expanding system \cite{Li:2008qm,Petersen:2009mz}. Here, we employ a hadron gas equation of state (HG) \cite{Zschiesche:2002zr} to investigate the differences in the dynamics and viscosity effects, a chiral equation of state with a moderate first order phase transition and a critical endpoint (CH) \cite{Papazoglou:1998vr,Zschiesche:2006rf,Steinheimer:2007iy} and a bag model equation of state with a rather large latent heat (BM) \cite{Rischke:1995mt}.

\section{Elliptic Flow Results}
\label{flow}

\begin{figure}[hb]
\centering  
\includegraphics[width=0.7\textwidth]{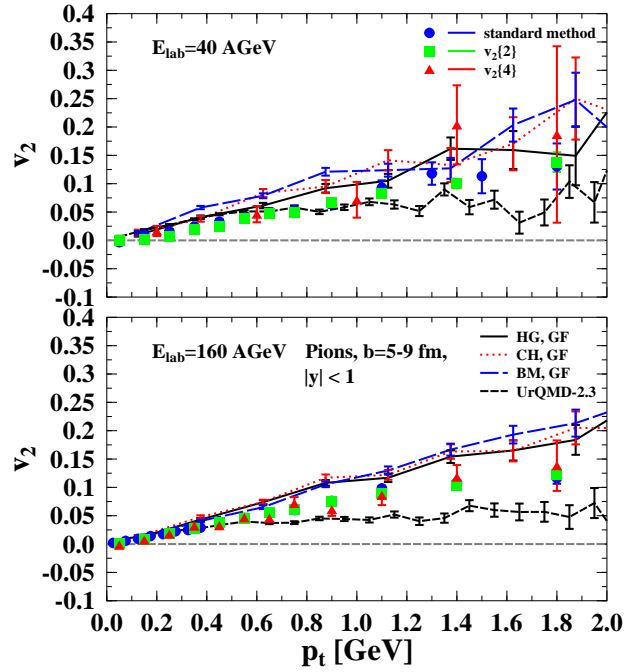}
\vspace{-1cm}
\caption{Elliptic flow of pions in mid-central (b=5-9 fm) Pb+Pb collisions at $E_{\rm lab}=40A~$GeV and  $E_{\rm lab}=160A~$GeV. The full black line depicts the hybrid model calculation with hadron gas EoS (HG), the red dotted line represents the chiral EoS (CH) and the blue long-dashed line the bag model EoS (BM) while the pure transport calculation is shown as the black dashed line (UrQMD-2.3). The colored symbols display experimental data obtained with different measurement methods by NA49 \cite{Alt:2003ab}. }
\label{v2_ptpi}      
\end{figure}

In Fig. \ref{v2_ptpi} the elliptic flow of pions as a function of transverse momentum, calculated within the pure transport approach (UrQMD-2.3) and in the hybrid approach employing different equations of state (HG,CH and BM), for mid-central (b=5-9 fm) Pb+Pb collisions at $E_{\rm lab}=40A$ GeV and $E_{\rm lab}=160A$ GeV is shown. The symbols represent the corresponding experimental data that have been measured by the NA49 collaboration applying different measurement methods \cite{Alt:2003ab}. The hydrodynamic evolution leads to higher elliptic flow values especially at higher $p_t$ even without an explicit phase transition, while the pure transport calculation underpredicts the data. Within this new hybrid approach, with more realistic initial conditions and freeze-out prescription, the effect of viscosity and non-equilibrium dynamics seems to be much stronger than the equation of state dependence.   

This finding is quite surprising since the expansion times differ for the three different scenarios as it can be estimated from the $R_O/R_S$ ratio \cite{Li:2008qm}. The overall magnitude of the pion transverse momentum is also affected by the EoS \cite{Petersen:2009mz}, but the anisotropy remains unchanged. From this, one could draw the conclusion that most of the elliptic flow is already built up during the initial scatterings in UrQMD, but we have checked that most of the flow is generated as expected during the hydrodynamic evolution. Therefore, we conclude that the effect of the softer EoS which should reduce the elliptic flow is cancelled by the longer expansion times in this case. 

\section{Summary and Conclusions}
Calculations of the transverse momentum dependence of elliptic flow have been presented for a integrated Boltzmann+hydrodynamics approach. This model is able to simulate heavy ion reactions from initial to the final state on an event-by-event-basis. The initial non-equilibrium dynamics and the final state rescatterings are taken care of by a transport approach while a hydrodynamic evolution is used for the hot and dense stage of the collision. This setup allows for a direct comparison between a transport approach and ideal fluid dynamics and, furthermore, the equation of state can be varied without adjusting the initial conditions and the freeze-out. The elliptic flow has been shown to be not sensitive to changes in the EoS while the smaller mean free path in the hydrodynamic evolution reflects directly in higher flow results which are consistent with the experimental data.

\section*{Acknowledgements}

We are grateful to the Center for the Scientific Computing (CSC) at Frankfurt for the computing resources. The authors thank Dirk Rischke for providing the 1 fluid hydrodynamics code. This work was supported by GSI and BMBF. This work was supported by the Hessian LOEWE initiative through the Helmholtz International Center for FAIR (HIC for FAIR).  


\end{document}